\documentclass[prd,nofootinbib,showpacs]{revtex4}
\usepackage{graphicx}
\usepackage{latexsym}
\usepackage{color}

\newcommand{\nn}{\nonumber}
\newcommand{\be}{\begin{equation}}
\newcommand{\ee}{\end{equation}}
\newcommand{\bea}{\begin{eqnarray}}
\newcommand{\eea}{\end{eqnarray}}

\begin{document}

\title{Cosmography beyond Standard Candles and Rulers}

\author{Jun-Qing Xia${}^1$, Vincenzo Vitagliano${}^{1,2}$, Stefano Liberati${}^{1,2}$, Matteo Viel${}^{2,3}$}

\affiliation{${}^1$Scuola Internazionale Superiore di Studi
Avanzati, Via Bonomea 265, 34136 Trieste, Italy}

\affiliation{${}^2$INFN sez. Trieste, Via Valerio 2, 34127 Trieste,
Italy}

\affiliation{${}^3$INAF-Osservatorio Astronomico di Trieste, Via
G.B. Tiepolo 11, I-34131 Trieste, Italy}

\begin{abstract}
  We perform a cosmographic analysis using several cosmological
  observables such as the luminosity distance moduli, the volume
  distance, the angular diameter distance and the Hubble parameter.
  These quantities are determined using different data sets: Supernovae
  type Ia and Gamma Ray Bursts, the Baryonic Acoustic Oscillations,
  the cosmic microwave background power spectrum and the Hubble
  parameter as measured from surveys of galaxies. This data set allows
  to put constraints on the cosmographic expansion with unprecedented
  precision.  We also present forecasts for the coefficients of the
  kinematic expansion using future but realistic data sets:
  constraints on the coefficients of the expansions are likely to
  improve by a factor ten with the upcoming large scale structure
  probes. Finally, we derive the set of the cosmographic parameters
  for several cosmological models (including $\Lambda$CDM) and compare
  them with our best fit set. While distance measurements are unable
  to discriminate among these models, we show that the inclusion of
  the Hubble data set leads to strong constraints on the lowest order
  coefficients and in particular it is incompatible with $\Lambda$CDM
  at 3-$\sigma$ confidence level. We discuss the reliability of this
  determination and suggest further observations which might be of
  crucial importance for the viability of cosmographic tests in the
  next future.
\end{abstract}


\maketitle


\section{Introduction}

Cosmology is going through a golden age as we can nowadays start
reconstructing the expansion history of the universe with
unprecedented precision.  The huge range of data sets spans a wide
realm of observations with heterogeneous nature, providing us with a
much more accurate tool for investigating the evolution of the
universe \cite{exp}. In particular, most of the observations agree
on the evidence that the universe is undergoing an era of positively
accelerated expansion, requiring the existence of a (more or less
conservative) source able to produce it. Since a whole set of
cosmological models explaining the late time acceleration has been
investigated, it goes without saying that a sensible test to
discriminate among different cosmological evolutions is related to a
proper interpretation of high redshift data.

Among these proposed models an important role is played by
extended/modified theories of gravity, i.e. by models in which the
late time acceleration is a by-product of some modified
gravitational dynamics (see Ref.\cite{ModifiedTheories} for a
review). In this context, it is clear that the development of a
``gravitational dynamics independent'' reconstruction of the
expansion history of our universe does play a crucial role.

Cosmography provides such unbiased test of the cosmological history
by assuming just homogeneity and isotropy and then use the so
obtained Friedman-Lema\^{\i}tre-Robertson-Walker (FLRW) metric to
express the distances\footnote{Note that depending on which physical
quantity one is measuring, it could be more convenient to extract
from some data set a particular distance indicator than another one.
These different quantities have different expressions of the Taylor
expansion in redshift, such that it could be more natural to
estimate cosmographic parameters, whose expression instead does not
depend on the analytic expansion, in one of these particular
frameworks. This ambiguity led also to a misconception about the
appropriate definition of distance one should investigate. From now
on we will refer only to: luminosity distance as the most direct
choice in the case of measures of distance for Supernovae Type Ia
(SNeIa) and Gamma Ray Bursts (GRBs); volume distance for Baryon
Acoustic Oscillations (BAOs); angular diameter distance for the
Cosmic Microwave Background (CMB) (see below for their definitions).
For a critical discussion about such difficulties, see
Ref.\cite{Cattoen:2008th}.} of the observed objects as power series
in a suitable redshift parameter. The coefficients of such powers,
casted into a combination of successive weighted derivatives of the
scale factor $a(t)$, contain the relevant information for a
kinematic description of the universe \cite{Cattoen:2008th, chiba,
sahni, vc_taylor, Rapetti:2006fv, viss1, viss2}.

A comment is necessary here: as already stressed in
Ref.\cite{vc_taylor} the ill-behavior at high $z$ (close and higher
than $z\approx 1$) of the usual redshift expansions strongly affects
the results leading in general to an underestimate of the errors. In
order to avoid these problems, as well as to control properly the
approximation associated with the truncation of the expansion, it is
useful to recast all the involved quantities as functions of an
improved parameter $y = z/(1 + z)$ \cite{vc_taylor, CPL,
Vitagliano:2009et}. In such a way, being $z \in (0,\infty)$ mapped
into $y\in(0, 1)$, it becomes possible to retrieve {improved
convergence properties of the Taylor series at high redshift
\cite{vc_taylor, viss3}.}

Some recent papers handle the problem of interpreting the data under
a cosmographic perspective using different probes
\cite{Vitagliano:2009et, capoz1, capoz2, CMB}. In this paper we are
going to explore the whole ensemble of data sets and use it to
constrain the parameters appearing in the expansions of the
characteristic scales associated to these indicators: Supernovae and
Gamma Ray Bursts, Baryon Acoustic Oscillations, Cosmic Microwave
Background and Hubble parameter estimates (Hub).  It is quite
obvious that by adding higher order powers to the redshift
expansions of such scales it is possible to improve the data
fitting, since more free parameters are involved. However, for a
given data set, there will be an upper bound on the order which is
statistically significant in the data analysis
\cite{Vitagliano:2009et}. In this sense it is crucial to always
determine the order of the expansion which maximizes the statistical
significance of the fit for a given data set or an ensemble of them.
We shall hence determine such order by performing suitable $F$-tests
depending on the collection of data sets we shall consider.
{However, we will show that these statistical arguments can be
misleading whenever is the case of a model, favored by the $F$-test,
but with error intervals so large to make the last determined
parameter completely unconstrained.} The parameters {obtained within
the cosmographic analysis} will then finally allow to evaluate, in a
dynamic independent way, the viability of any theory aiming to
explain the current expansion of the universe.

The plan of the paper is as follows: in Section \ref{s2} we will
describe the cosmographic expansion and the theoretical framework
(also partly reported in the Appendix), while in Section \ref{s3} we
will present the observational probes used for the present analysis.
Section \ref{s4} contains the main results using the fourth and
fifth order expansions. In Section \ref{s5} we focus on the
forecasts with futuristic data sets {while Section \ref{s6} is
devoted to the study and comparison of the cosmographic expansions
for different cosmological models.} We conclude in Section \ref{s7}.

\section{Cosmographic expansions}
\label{s2}

As a pedagogical example, we will discuss first the procedure that
has been followed in order to obtain the cosmographic expansion for
the luminosity distance. As already pointed out, we will start from
the only assumption that the universe is homogeneous and isotropic,
so that the metric describing its properties is the FLRW one \be
ds^2=-c^2dt^2+a^2(t)\left[\frac{dr^2}{1-kr^2}+r^2d\Omega^2\right]\,;
\ee using this metric, it is possible to express the luminosity
distance $d_L$ as a power expansion in the redshift parameter $z$
(or in term of the $y$-parameter), where the coefficients of the
expansion are some functions of the scale factor $a(t)$ and its
higher order derivatives.

Following Ref.\cite{wei},  the relation between the apparent
luminosity $l$ of an object and its absolute luminosity $L$ defines
the luminosity distance $d_L$ \bea l=\frac{L}{4 \pi r^2_1
a^2(t_0)(1+z)^2}= \frac{L}{4\pi d_L^2}\,, \eea where $r_1$ is the
comoving radius of the light source emitting at time $t_1$, $t_0$ is
the later time an observer in $r=0$ is catching the photons, and
redshift $z$ is, as usual, defined as $1+z=a(t_0)/a(t_1)$. The
radial coordinate $r_1$ in a FLRW universe can be written for small
distances as~\cite{viss1} \be
r_1=\int_{t_1}^{t_0}\frac{c}{a(t)}dt-\frac{k}{3!}\left[\int_{t_1}^{t_0}\frac{c}{a(t)}dt\right]^3+\mathcal{O}\left(\left[\int_{t_1}^{t_0}\frac{c}{a(t)}dt\right]^5\right)\,,
\ee with $k=-1,~0,~+1$ respectively for hyperspherical, Euclidean or
spherical universe. In such a way, it is possible to recover the
expansion of $d_L$ for small $z$ \bea
d_L(z)=cH_0^{-1}\!\left\{z+\frac{1}{2}(1-q_0)z^2-\frac{1}{6}\left(1-q_0-3q_0^2+j_0+
\frac{kc^2}{H_0^2a^2(t_0)}\right)z^3+\mathcal{O}(z^4)\right\},\nn\\
\eea having defined the cosmographic parameters as \bea
H_0&\equiv&\left.\frac{1}{a(t)}\frac{da(t)}{dt}\right|_{t=t_0}\equiv\left.\frac{\dot{a}(t)}{a(t)}\right|_{t=t_0}~,\nn\\
q_0&\equiv&-\left.\frac{1}{H^2}\frac{1}{a(t)}\frac{d^2a(t)}{dt^2}\right|_{t=t_0}\equiv\left.-\frac{1}{H^2}\frac{\ddot{a}(t)}{a(t)}\right|_{t=t_0}~,\nn\\
j_0&\equiv&\left.\frac{1}{H^3}\frac{1}{a(t)}\frac{d^3a(t)}{dt^3}\right|_{t=t_0}\equiv\left.\frac{1}{H^3}\frac{a^{(3)}(t)}{a(t)}\right|_{t=t_0}~.
\eea
If instead we use the redshift variable $y=z/(1+z)$, the definition
of the cosmographic parameters will not be affected, while now the
luminosity distance turns out to be \bea
d_L(y)=\frac{c}{H_0}\Bigg\{y-\frac{1}{2}(q_0-3)y^2+\frac{1}{6}\left[11-5q_0+3q^2_0-j_0+\Omega_{k_0}\right]y^3+\mathcal{O}(y^4)\Bigg\}~,\nn\\
\eea where $\Omega_{k_0}=-kc^2/H_0^2a^2(t_0)$ is the spatial
curvature energy density.  For a flat universe, $\Omega_{k_0}=0$.
Since we are interested in spanning the universe at any redshift, in
the following we will use only the formulation of the expansion in
the variable $y$. In our analysis we will put constraints up to
the cosmographic parameters $s_0$ and $c_0$, namely:
\bea
s_0&\equiv&\left.\frac{1}{H^4}\frac{1}{a(t)}\frac{d^4a(t)}{dt^4}\right|_{t=t_0}\equiv\left.\frac{1}{H^4}\frac{a^{(4)}(t)}{a(t)}\right|_{t=t_0}~,\nn\\
c_0&\equiv&\left.\frac{1}{H^5}\frac{1}{a(t)}\frac{d^5a(t)}{dt^5}\right|_{t=t_0}\equiv\left.\frac{1}{H^5}\frac{a^{(5)}(t)}{a(t)}\right|_{t=t_0}~.
\eea In the Appendix we list the cosmographic series for the
physical quantities involved in our analysis.

\section{Observational data sets}
\label{s3}

In our calculations, we rely on the following current observational
data sets: i) SNIa and GRB distance moduli;
ii) BAO in the galaxy power spectra; iii) CMB; iv) Hubble parameter determinations.

The SNIa distance moduli provide the luminosity distance as a
function of redshift $D_{\rm L}(z)$. In this paper we will use the
latest SNIa data sets from the Supernova Cosmology Project, ``Union2
Compilation'' which consists of 557 samples and spans the redshift
range $0\lesssim z \lesssim 1.55$ \cite{Amanullah:2010vv}.  In this
data set, they improved the data analysis method by using and
refining the approach of their previous work \cite{Kowalski:2008ez}.
When comparing with the previous ``Union Compilation'', they
extended the sample with the supernovae from
Refs.\cite{Amanullah:2010vv,Amanullah:2007yv}.  The authors also
provide the covariance matrix of data with and without systematic
errors and, in order to be conservative, we include systematic
errors in our calculations.

In addition, we also consider another luminosity distance indicator
provided by GRBs, that can potentially be used to measure the
luminosity distance out to higher redshift than SNIa. GRBs are not
standard candles since their isotropic equivalent energetics and
luminosities span $3-4$ orders of magnitude. However, similarly to
SNIa it has been proposed to use correlations between various
properties of the prompt emission and also of the afterglow emission
to standardize GRB energetics (e.g. Ref.\cite{Ghirlanda:2004fs}).
Recently, several empirical correlations between GRB observables
were reported, and these findings have triggered intensive studies
on the possibility of using GRBs as cosmological ``standard''
candles. However, due to the lack of low-redshift long GRB data to
calibrate these relations, in a cosmology-independent way, the
parameters of the reported correlations are given assuming an input
cosmology and obviously depend on the same cosmological parameters
that we would like to constrain. Thus, applying such relations to
constrain cosmological parameters leads to biased results. In
Ref.\cite{Li:2007re} this ``circular problem'' is naturally
eliminated by marginalizing over the free parameters involved in the
correlations; in addition, some results show that these correlations
do not change significantly for a wide range of cosmological
parameters \cite{Firmani:2006hq,Schaefer:2006pa}. Therefore, in this
paper we use the 69 GRBs over a redshift range $z \in [0.17, 6.60]$
presented in Ref.\cite{Schaefer:2006pa}, but we keep into account in
our statistical analysis the issues related to the circular problem
that are more extensively discussed in Ref.\cite{Li:2007re} and also
the fact that all the correlations used to standardize GRBs have
scatter and a poorly understood physics. For a more extensive
discussion and for a full presentation of a GRB Hubble Diagram with
the same sample that we used we refer the reader to section 4 of
Ref.\cite{Schaefer:2006pa}.

In the calculation of the likelihood from SNIa and GRBs, we have
marginalized over the absolute magnitude $M$ which is a nuisance
parameter, as done in Refs.\cite{Goliath:2001af,DiPietro:2002cz}
\begin{equation}
\bar{\chi}^2=A-\frac{B^2}{C}+\ln\left(\frac{C}{2\pi}\right)~,
\end{equation}
where
\begin{eqnarray}
&&A=\sum_i\frac{(\mu^{\rm data}-\mu^{\rm th})^2}{\sigma^2_i}~,~
B=\sum_i\frac{\mu^{\rm data}-\mu^{\rm
th}}{\sigma^2_i}~,~C=\sum_i\frac{1}{\sigma^2_i}~.
\end{eqnarray}

BAOs have been detected in the current galaxy redshift survey data
from the SDSS and the Two-degree Field Galaxy Redshift Survey
(2dFGRS) \cite{Eisenstein:2005su,Cole:2005sx,Percival:2009xn}.  They
can directly measure not only the angular diameter distance, $D_{\rm
A}(z)$, but also the expansion rate of the universe, $H(z)$, a
powerful tool for studying dark energy \cite{Albrecht:2006um}.
Since current BAO data are not accurate enough for extracting the
information of $D_{\rm A}(z)$ and $H(z)$ separately
\cite{Okumura:2007br}, one can only determine an effective
``volume'' distance \cite{Eisenstein:2005su}
\begin{equation}
D_{\rm V}(z)\equiv\left[(1+z)^2D_{\rm
A}^2(z)\frac{cz}{H(z)}\right]^{1/3}~.
\end{equation}

In this paper we use the Gaussian priors on the distance ratio of
the volume distances as recently extracted from the SDSS and 2dFGRS
surveys \cite{Percival:2009xn} at $z=0.35$ and at $z=0.2$ (the two
mean redshifts of the surveys)
\begin{equation}\label{bao}
\frac{D_{\rm V}(z=0.35)}{D_{\rm
V}(z=0.2)}=1.736\pm0.065~(1\,\sigma)~.
\end{equation}
The $\chi^2$ of BAO data used in the Monte Carlo Markov Chain
analysis will thus be
\begin{equation}
\chi^2_{\rm BAO}=\frac{(D_{\rm V}(z=0.35)/D_{\rm
V}(z=0.2)-1.736)^2}{0.065^2}~.
\end{equation}

It is worth stressing here that actually both the physics and the
data of BAOs depend on the content in matter of the universe
$\Omega_{\rm m}$. Hence, they are {\em a priori} dependent on a
chosen dynamical framework (see also Ref.\cite{bass} for a review).
This issue is usually ignored in the data analyses performed in the
literature. However, such an approximation turns out to be valid if
one does not range far away from the typical fiducial model firstly
used in the determination of the physical data points. Indeed, the
deviation of different models from the fiducial one can be
parametrized and estimated by the ratio $D_V(\textrm{new
model})/D_V(\textrm{fiducial model})$. The impact of the spacetime
priors on the power spectrum measurement was analyzed in ref.
\cite{teg} and led to the conclusion that the ratio Eq.(\ref{bao})
is only weakly dependent on dynamical features.  Hence, we can
safely use BAOs as a further tool to constrain the cosmographic
parameters.

Next step in our analysis is the inclusion of the CMB measurement
which is sensitive to the distance from the last scattering surface
via the locations of peaks and troughs of the acoustic oscillations.
This data constrains the curve of the cosmological history at very
high redshift, $z\simeq1100$, and hence could be very helpful for
discriminating among competing theoretical models for dark energy,
as they necessarily have to coincide at $z\leq 1$ -- see for example
ref. \cite{ede}. The sound horizon at decoupling\footnote{In our
calculation, we choose $z_\ast=1091.3$, the best fit value obtained
by the WMAP group \cite{WMAP7}.}, $r_s(z_\ast)$, sets a physical
scale for the baryon-photon oscillations depending on the baryon
density, the photon energy density, and the cold dark matter
density. Now, it is known that the angular diameter distance
$D_A(z)$ describes the ratio between the proper size of an object at
a certain redshift $z$ and the correlated observed angular size. The
angle $\theta_A$ under which the sound horizon is observed today is
given by \be\label{teta} \theta_A\equiv\pi l_A^{-1}\equiv
r_s(z_\ast)/D_A(z_\ast)=0.593^\circ\pm0.001^\circ~(1\,\sigma)\,, \ee
where $l_A$ denotes the location of the first peak in the multipole
space. As for BAOs, the dependence on the cosmological density
parameters would not allow the use of CMB observables in a fully
cosmographic approach. However, following Ref.\cite{rasa} is
possible to give model-independent cosmological constraints if one
specifies some extra physical assumptions to be fullfilled by
cosmological models. The CMB power spectrum today (apart from the
low multipoles shape) is shared by models having the same primordial
perturbation spectra and the same value of $\Omega_{\rm CDM}$ and
$\Omega_{\rm baryon}$. For this reason a basically model-independent
approach can be pursued by restricting our analysis to models having
a standard physics up to the decoupling era; asking that new physics
after decoupling only modifies the small angle spectrum by changing
the overall amplitude and $D_A(z_\ast)$, and requiring that any
multipole-dependent effect at late time remain small. Such
assumptions, while is cutting away some models like $f(R)$ models
with no Dark Matter \cite{noDM} or models with new radiation degrees
of freedom, are still general enough to cover most of the
cosmological models on the market.

Finally, we add the direct
determinations of the Hubble parameter $H(z)$ to constrain the
cosmographic expansion. Since the Hubble parameter depends on the
differential age of the Universe as a function of redshift,
\begin{equation}
H(z)=-\frac{1}{1+z}\frac{dz}{dt}~,
\end{equation}
measuring the $dz/dt$ could directly estimate $H(z)$.
Ref.\cite{Jimenez:2003iv} used the Sloan Digital Sky Survey data and
obtained a measurement of $H(z)$ at the redshift $z\simeq0$. In
Ref.\cite{Simon:2004tf}, the public data of Gemini Deep Survey
(GDDS) survey \cite{Abraham:2004ra} and archival data
\cite{Treu:1999aa} were used in order to get the differential ages
of galaxies. In practice, they selected samples of passively
evolving galaxies with high-quality spectroscopy, and then used
stellar population models to constrain the age of the oldest stars
in these galaxies (we refer to their paper for a more exhaustive
explanation of the method used). After that, they computed
differential ages at different redshift bins and obtained eight
determinations of the Hubble parameter $H(z)$ in the redshift range
$z\in[0.1,1.8]$. We calculate the $\chi^2$ value of this Hubble
parameter data by using
\begin{equation}
\chi^2_{\rm Hub}=\sum^{9}_{i=1}\frac{(H^{\rm th}(z_i)-H^{\rm
obs}(z_i))^2}{\sigma^2_{H}(z_i)}~,
\end{equation}
where $H^{\rm th}(z)$ and $H^{\rm obs}(z)$ are the theoretical and
observed values of Hubble parameter, and $\sigma_{H}$ denotes the
error bar of observed data. We also make use of the newly released
prior on the Hubble parameter $H_0$, which consists of  a
measurement of the Hubble parameter obtained by the Near Infrared
Camera and Multi-Object Spectrometer (NICMOS) Camera 2 of the Hubble
Space Telescope (HST).

These observations fix the parameter $H_0 = 100h_0\,{\rm
(km/s)}/{\rm Mpc}$ by a Gaussian likelihood function centered around
$H_0 = 74.2\,{\rm (km/s)}/{\rm Mpc}$ and with a standard deviation
$\sigma = 3.8 \,{\rm (km/s)}/{\rm Mpc}$ \cite{Riess:2009pu}. We
stress that all the mentioned methods for determining $H(z)$ are basically
``gravitation theory independent''.

An important point must be underlined: the Taylor series of the
Hubble parameter already includes into the coefficient of the $n$-th
$y$-power the same number of cosmographic parameters of the other
series expanded up to the $(n+1)$-th $y$-power (see Appendix). This
is due, in comparison with the other distance definitions above, to
an extra derivative with respect to time included in the definition
of the Hubble parameter (see also Ref.\cite{CMB}).

{ For this reason, and for the different nature of the Hubble data,
  we will initially consider constraints (based on standard candles
  and rulers) of the expansion coefficients associated to different
  notions of distances; at the end, we will add the Hubble data using
  one order less in the $y$-power expansion with respect to the
  distance data in order to constrain the same set of parameters.}

In order to compute the likelihood, we use a Monte Carlo Markov
Chain technique as it is usually done in order to explore
efficiently a multi-dimensional parameter space in a Bayesian
framework. For each Monte Carlo Markov Chain calculation, we run
four independent chains that consist of about 300,000 $-$ 500,000
chain elements each. We test the convergence of the chains by using
the Gelman and Rubin criterion \cite{MCMC} with $R-1$ of order 0.01,
which is more conservative than the often used and recommended value
$R-1 < 0.1$ for standard cosmological calculations.

\begin{table*}
\caption{Constraints on the cosmography parameters up to fifth order
term from different data combinations.}\label{table:4th}
\begin{center}
\begin{tabular}{c|ccccc}
  \hline
  \hline
  Data&\multicolumn{5}{c}{SNIa}\\
  \hline
  Parameter&$q_0$&$j_0$&$s_0$&$c_0$&$H_0$\\
  \hline
   Best Fit&$-0.41$&$-1.99$&$-$&$-$&$-$\\
  Mean&$-0.41\pm0.16$&$-1.99\pm1.36$&$-$&$-$&$-$\\
  $\chi^2_{\rm min}/$d.o.f.&\multicolumn{5}{c}{549.69/555}\\
  \hline
  \hline
    Data&\multicolumn{5}{c}{SNIa+GRB}\\
  \hline
  Parameter&$q_0$&$j_0$&$s_0$&$c_0$&$H_0$\\
  \hline
  Best Fit&$-0.78$&$5.03$&$50.18$&$-$&$-$\\
  Mean&$-0.76\pm0.26$&$4.82\pm4.07$&$53.57\pm46.38$&$-$&$-$\\
  $\chi^2_{\rm min}/$d.o.f.&\multicolumn{5}{c}{628.70/623}\\
  \hline
  \hline
    Data&\multicolumn{5}{c}{SNIa+GRB+BAO+CMB $4^{th}$ order}\\
  \hline
  Parameter&$q_0$&$j_0$&$s_0$&$c_0$&$H_0$\\
  \hline
  Best Fit&$-0.32$&$-2.57$&$-18.40$&$-$&$-$\\
  Mean&$-0.28\pm0.17$&$-2.88\pm1.64$&$-17.61\pm2.56$&$-$&$-$\\
  $\chi^2_{\rm min}/$d.o.f.&\multicolumn{5}{c}{633.33 / 625}\\
  \hline
  \hline
    Data&\multicolumn{5}{c}{SNIa+GRB+BAO+CMB $5^{th}$ order}\\
  \hline
  Parameter&$q_0$&$j_0$&$s_0$&$c_0$&$H_0$\\
  \hline
  Best Fit&$-0.17$&$-6.92$&$-74.18$&$-10.58$&$-$\\
  Mean&$-0.49\pm0.29$&$-0.50\pm4.74$&$-9.31\pm42.96$&$126.67\pm190.15$&$-$\\
  $\chi^2_{\rm min}/$d.o.f.&\multicolumn{5}{c}{627.61/624}\\
  \hline
  \hline
    Data&\multicolumn{5}{c}{SNIa+GRB+BAO+Hub+CMB ($5^{th}$ order) +Hub ($4^{th}$ order)}\\
  \hline
  Parameter&$q_0$&$j_0$&$s_0$&$c_0$&$H_0$\\
  \hline
  Best Fit&$-0.24$&$-4.82$&$-47.87$&$-49.08$&$71.65$\\
  Mean&$-0.30\pm0.16$&$-4.62\pm1.74$&$-41.05\pm20.90$&$-3.50\pm105.37$&$71.16\pm3.08$\\
  $\chi^2_{\rm min}/$d.o.f.&\multicolumn{5}{c}{639.81/633}\\
  \hline
  \hline
\end{tabular}
\end{center}
\end{table*}

\section{Data analysis}
\label{s4}

In this section we present our main results on the constraints for
the cosmographic expansion from the current observational data sets.

With the accumulations of new data and the improvements of their
quality, it is of great interest to estimate the free parameters in
the polynomial terms of highest order. In our previous paper
\cite{Vitagliano:2009et} we already showed the inconsistency of the
results in the analysis of the cosmographic expansion caused by
early truncations of the power series. For these reasons we will now
present the results obtained for the most meaningful term of the
expansion. In order to find out which is the most viable truncation
of the series for a given data set, one can use a test comparing two
nested models (in this case, two different truncations of the Taylor
series). The $F$-test provides exactly this criterion of comparison,
identifying which of two alternatives fits better, and in the more
statistically significant way, the data.

In such test, one assumes the correctness of one of the models (the
one with less parameters), and then assess the probability for the
alternative model to fit the data as well. If this probability is
high, then no statistical benefit comes from the extra degrees of
freedom associated to the new model. Hence, the smaller the
probability, the more significant the data fitting of the second
model against the first one will be.  Quantitatively, the $F$-ratio
among the two polynomials is defined as \be
F\equiv\frac{(\chi_1^{\,2}-\chi_2^{\,2})}{\chi_2^{\,2}}~\frac{N-n_2}{n_2-n_1}\,,
\ee where $N$ is the number of data points, and $n_i$ represent the
number of parameters of the $i$-model. The $P$-value, \emph{i.e.}
the area subtended by the $F$-distribution curve delimited from the
$F$-ratio point, quantifies the viability of matching models as
already mentioned. We use the threshold of $5\%$ as the significance
level on the $P$-value under which the model with one more parameter
fits the data better than the other one.

We already found in Ref.\cite{Vitagliano:2009et} that variations of
the total energy density of the universe $\Omega_0=1-\Omega_{k_0}$,
with the spatial curvature parameter ranging between -1 and 1, have
a negligible effect on the cosmographic constraints. This is
basically due to the fact that the error bars for the cosmographic
parameters are still quite large in comparison with the best fit
values. Nonetheless, it is worth noting that this will not be
necessarily the case when future data, especially at moderate or
high redshift,  will substantially improve the constraints. It is
then possible that future cosmographic analysis might have to
include the spatial curvature effects in the reconstruction of the
overall history of the universe. This would be the cosmographic
expansion counterpart of the strong sensitivity on $\Omega_{k_0}$
showed by the reconstruction of $w(z)$ \cite{bass2}.

We then assume $\Omega_{k_0}=0$ in our analysis and only present the
results for the cosmographic parameters, instead of their
combinations with $\Omega_{k_0}$, since the effect of curvature can
be for the moment safely neglected. Table \ref{table:4th} shows the constraints on
the cosmography parameters as obtained from different data
combinations.

We start performing the data analysis with the SNIa data only. We
find that already at the fourth order term in the expansion, the
minimal $\chi^2$ is $549.59$. This is not reduced significantly when
compared with the constraint of the third order case, which has
$\chi^2_{\rm min}=549.69$. Hence, introducing the snap free
parameter $s_0$ does not improve the constraints. Indeed, using the
$F$-test, we find a $F$-ratio of 0.11 and a $P$-value of 73.93\%.
Therefore, cosmography up to the fourth order term does not fit the
SNIa data significantly better: the cosmographic expansion up to the
jerk term $j_0$ (third order) is enough.

\begin{figure}[t]
\begin{center}
\includegraphics[scale=0.26]{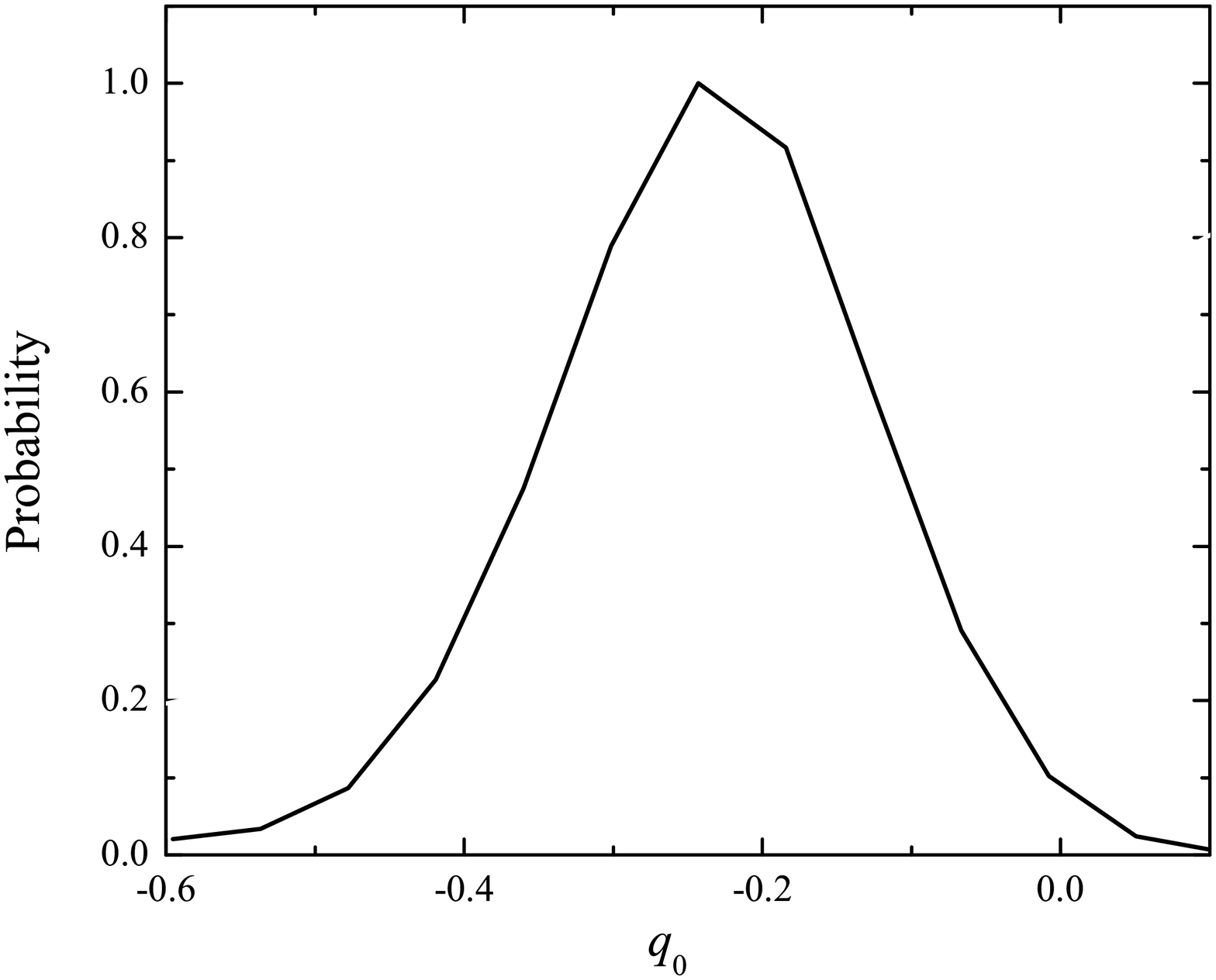}
\includegraphics[scale=0.26]{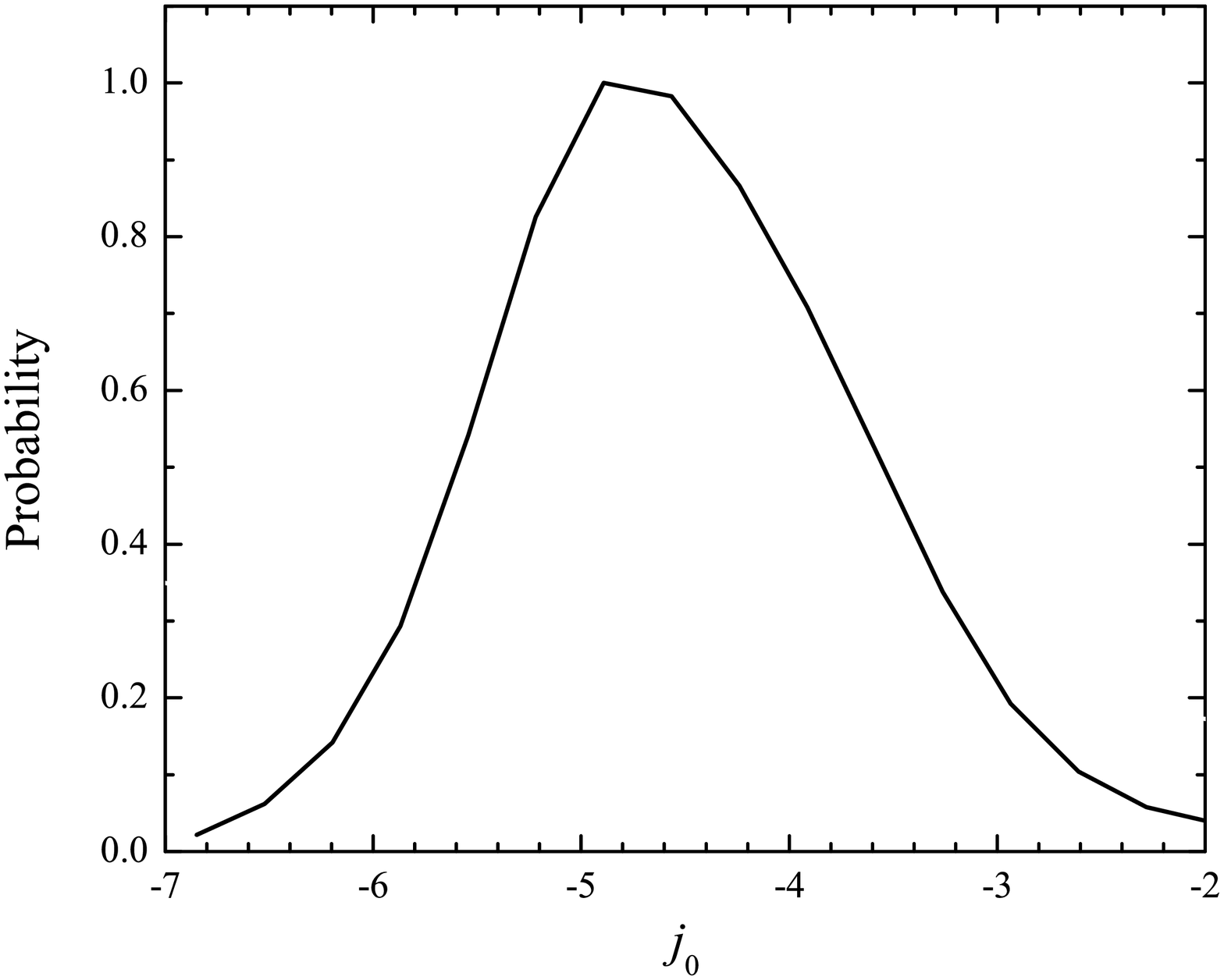}
\includegraphics[scale=0.26]{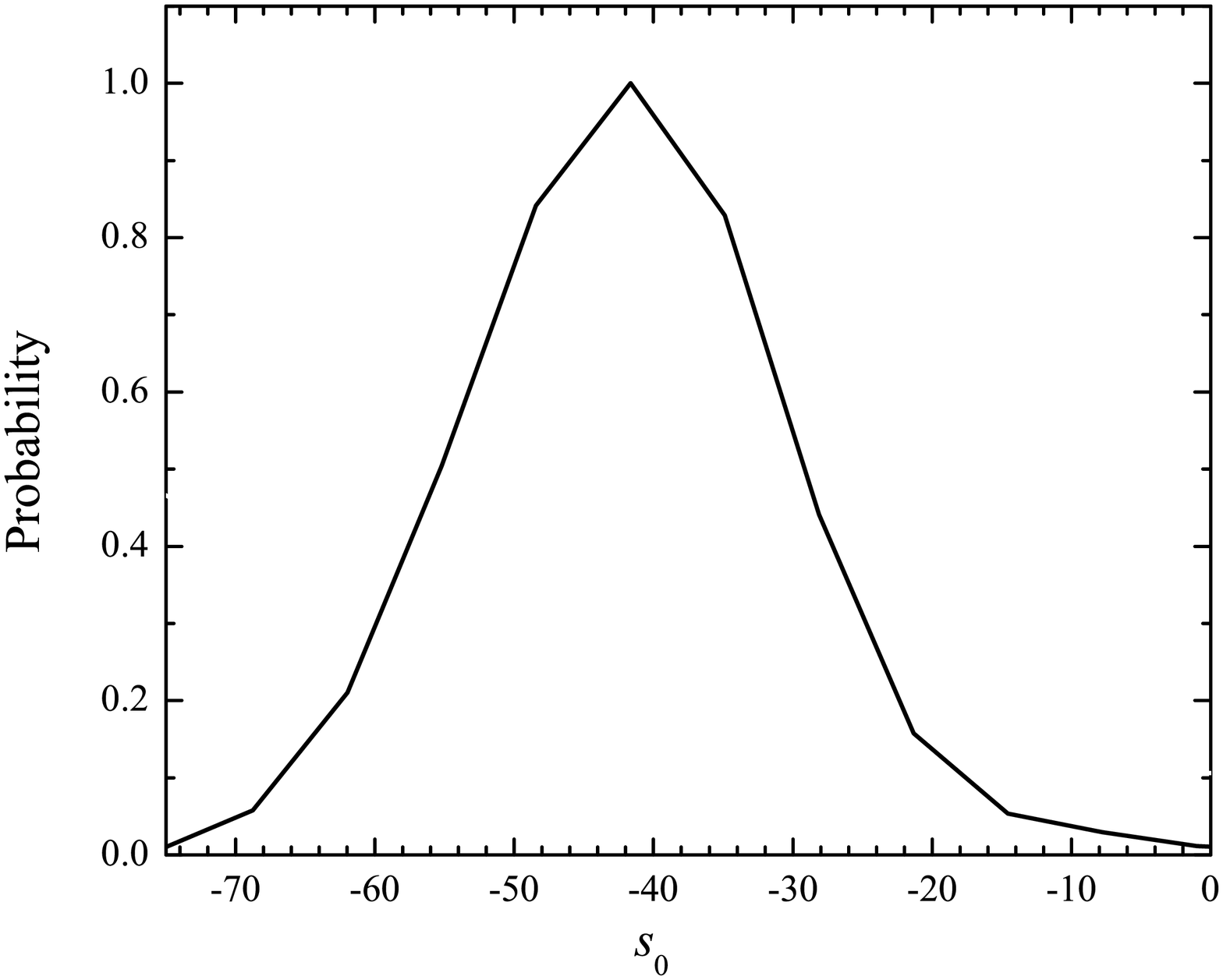}
\includegraphics[scale=0.26]{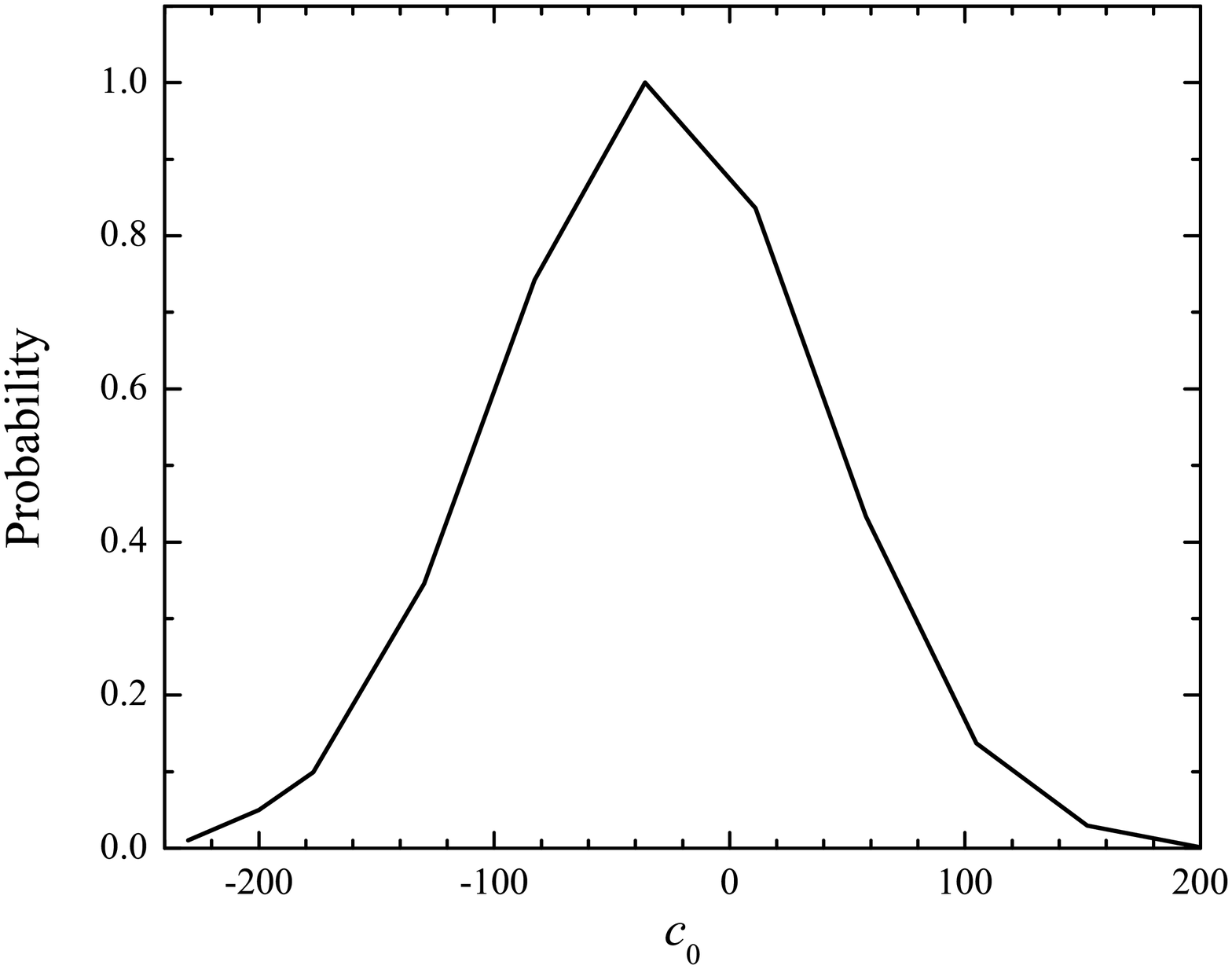}
\caption{One-dimensional likelihood distributions for the parameters
$q_0$, $j_0$, $s_0$ and $c_0$ for the data combinations SNIa+GRB+BAO+CMB+Hub. \label{fig:4th:1d}}
\end{center}
\end{figure}

After adding the GRB data, the fourth order case could give a better
constraint than third order only.  When comparing the SNIa+GRB
results, the minimal $\chi^2$ has been reduced by about five
($\chi^2=628.70$ instead of $\chi^2=633.32$). Using one more time
the $F$-test to contrast third and fourth order expansions, we find
$F$-ratio~$=4.59$, $P$-value~$=3.26\%$. Thus in this case the fourth
order term indeed helps to fit the observed data significantly
better. The inclusion of GRBs was found to constrain the
deceleration parameter $q_0$ as $q_0=-0.76\pm0.26$, so  confirming
that our universe is undergoing an accelerated expansion with a
confidence level which is marginally at $3\,\sigma$
\cite{Vitagliano:2009et}\footnote{Note that our best fit here is
different from the one reported in Ref.\cite{Vitagliano:2009et} due
to our use of the improved SN catalogue ``Union2 Compilation''.}.
The $F$-test does not suggest to further improve the expansion up to
fifth order.
%
%

Including the data point related to BAO does not improve
significantly the constraints. The constraining power of BAO is
rather weak since there is only one BAO data point and its redshift
is much lower than those of SN and GRB data. {For this reason we
will consider directly the data set that includes both BAO and CMB.}

When the CMB angle $\theta_A$ defined in Eq.(\ref{teta}) is added
into our analysis, the cosmographic curve is constrained at very
high redshift, $z_\ast\simeq1100$ (namely $y\simeq1$). Even though
the CMB observable is providing just one data point, due to its high
redshift it is in principle very helpful for discriminating among
competing theoretical models producing late time accelerated
expansion, since these necessarily converge to the same cosmological
history at small $z$ (see for example Ref.\cite{ede}). The large
difference between the two $\chi^2_{\rm min}$ of the fourth and
fifth order expansion in powers of $y$ of the distances, implies
that the latter is the (statistically) more reliable parametrization
($F_{\rm ratio}=5.69$ and $P_{\rm value}=0.02\%$), giving a result
very close to the $\Lambda$CDM prediction. Sixth order expansions
does not give any substantial statistical improvement.

%

As already stated at the end of the previous section, Hubble data
must be added and analyzed cautiously, since they are inhomogeneous
with respect to the previous data sets both in nature and
mathematical handling. In Table I we present directly the results
for the constraints up to the $c_0$ cosmographic parameter, since
this truncation turns out to be strongly favored with respect to the
previous nested model ($F_{\rm ratio}=19.77$ and $P_{\rm
value}<0.01\%$).

\begin{figure}[t]
\begin{center}
\includegraphics[scale=0.26]{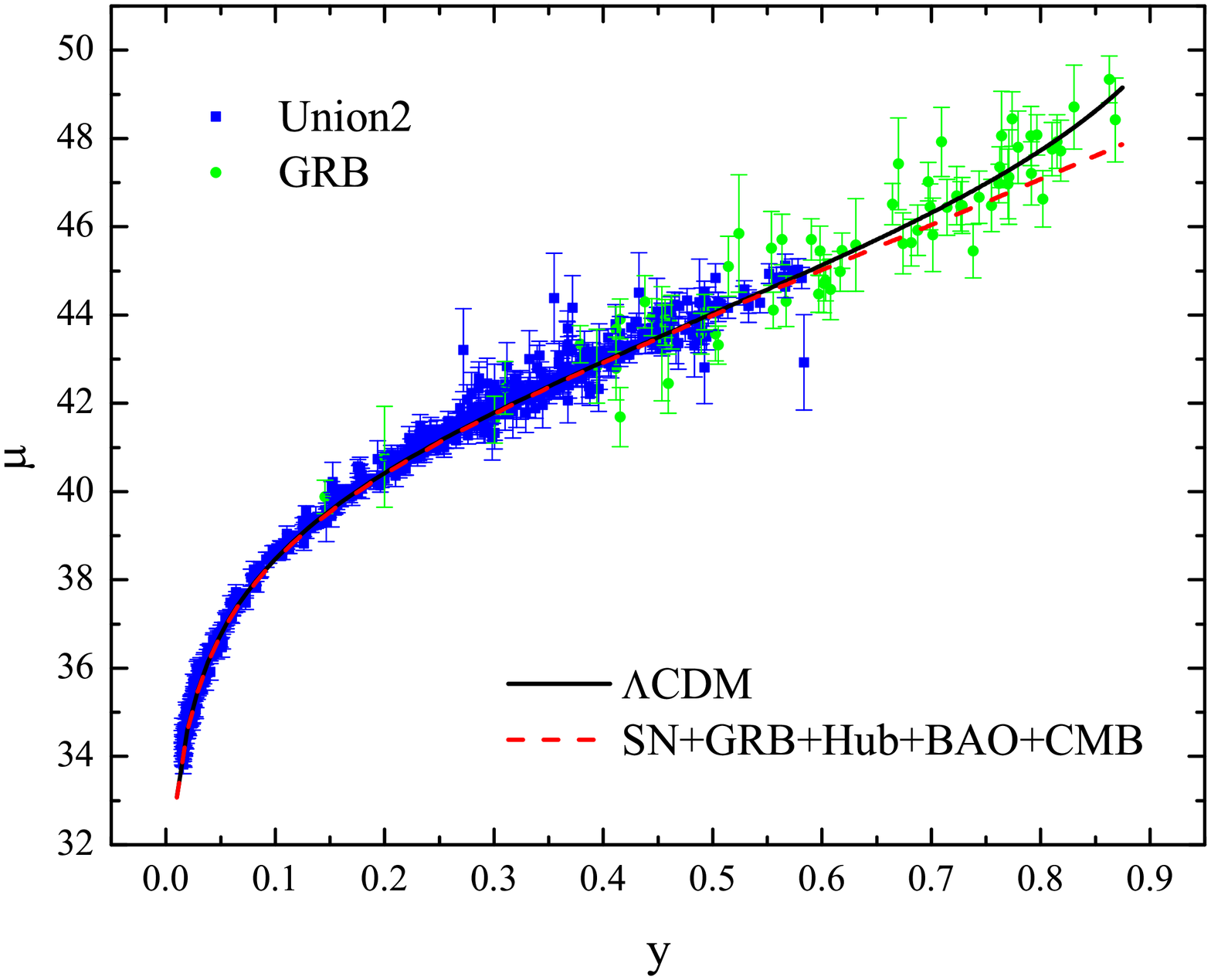}
\includegraphics[scale=0.26]{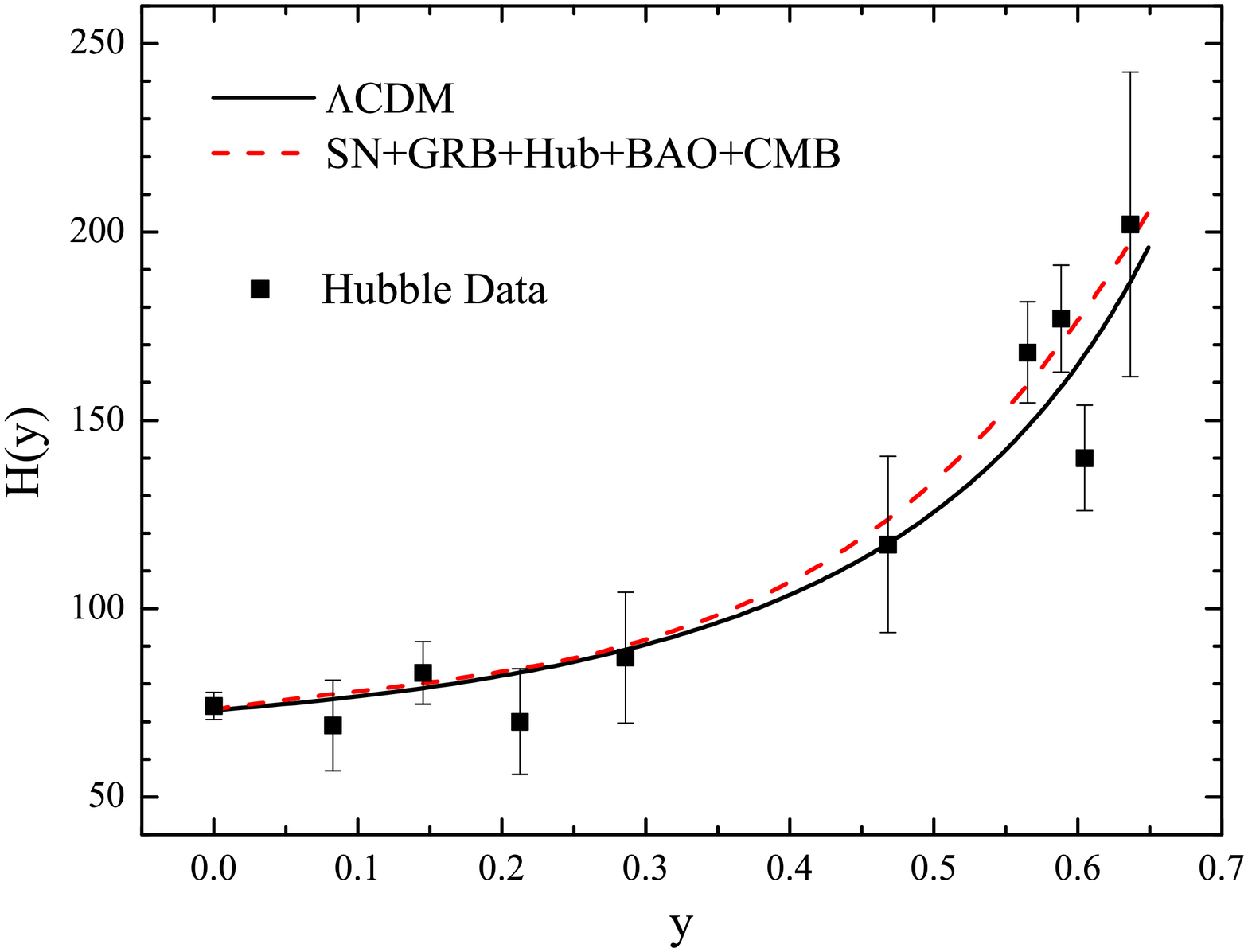}
\caption{Theoretical predictions of distance moduli (left panel) and
Hubble parameter (right panel) from the best fit model with
the full data combination, together
with the observed data sets. We also show the curves obtained in the
$\Lambda$CDM framework for comparison (thin black solid
lines).\label{fig:4th:th}}
\end{center}
\end{figure}

The theoretical curves of $\mu(z)$ and $H(z)$ are in good agreement
with the observed cosmological data, as shown in figure
\ref{fig:4th:th}. The constraint on $H_0$ is close to the usually
quoted value, namely at 68\% confidence level is $H_0=71.16\pm3.08$
(km/s)/Mpc. One can see that the addition of the Hubble data leads
to relevant improvements in the determination of the other
cosmographic parameters with the exception of $c_0$, which is still
basically unconstrained. We also checked whether the next
cosmographic parameter had to be included. We find that, for the
richest combination SN+GRB+BAO+CMB+Hub, the new $\chi^2_{\rm min}$,
is extremely close to the value in Table \ref{table:4th}. Therefore,
we stop our analysis here.

It is here interesting to underline the power of the $y$-expanded
series (convergent as long as $y<1$) allowing us to describe the
whole cosmological history with the use of relatively few
parameters. This circumstance becomes for example evident in the
left panel of figure \ref{fig:4th:th}, where the furthest GRB data
point reaches the distance of $y\simeq0.87$ (corresponding to
$z\simeq6.6$).
%

\section{Forecasting}
\label{s5}

Since the present data do not give yet very stringent constraints on
the parameters of cosmography, especially for the parameter of fifth
order term, it is worthwhile discussing whether future data could
determine these parameters more effectively. For this purpose in
what follows we shall perform new analysis of possible future
constraints by choosing, as a fiducial model, the best fit parameter
set for the cosmographic expansion up to the fifth order term as
fixed by the combination of all the previously considered data sets.

The projected satellite SNAP (Supernova / Acceleration Probe) would
be a space based telescope with a one square degree field of view
with $10^9$ pixels. It aims at increasing the discovery rate for
SNIa to about $2000$ per year in the redshift range $0.2 < z < 1.7$.
In this paper we simulate about $2000$ SNIa according to the
forecast distribution of the SNAP. For the error, we follow the
Ref.\cite{snap} which takes the magnitude dispersion $0.15$ and the
systematic error $\sigma_{\rm sys} = 0.02 \times z/1.7$. The whole
error for each data is given by $\sigma_{\rm
mag}(z_i)=\sqrt{\sigma^2_{\rm sys}(z_i)+0.15^2/n_i}$, where $n_i$ is
the number of Supernovae of the $i$-th redshift bin. Furthermore, we
add as an external data set a mock data set of 400 GRBs, in the
redshift range $0.4 < z < 6.4$ with an intrinsic dispersion in the
distance modulus of $\sigma_\mu = 0.16$ and with a redshift
distribution very similar to that of figure 1 of
Ref.\cite{Hooper:2007aa}.

Regarding a future BAO data set, we adopt the predicted performance
of the BOSS survey in SDSS III, which will measure the angular
diameter distance $d_A(z)$ and the Hubble expansion rate $H(z)$ of
the Universe over a broad range of redshifts. The measurement
precision for $d_A(z)$ is $1.0\%$, $1.1\%$, and $1.5\%$ at $z =
0.35$, $0.6$, and $2.5$, respectively, and the forecast precision
for the $H(z)$ is $1.8\%$, $1.7\%$, and $1.2\%$ at the same
redshifts \cite{BOSS}.  We also impose a Gaussian prior for the
current Hubble parameter $H_0$ with the error of $1\%$ provided by a
future direct measurement.

Next coming CMB measurement, mainly via the Planck satellite,
could give quite accurate constraints on the cosmological
parameters. The error bar of $\theta_A$ could be shrunk by a factor
of 3, namely, the standard derivation
$\sigma_{\theta_A}=0.0003^\circ$.

Using all these future mock data, we get the standard derivations of
cosmographic parameters: $\sigma_{q_0}=0.02$,
$\sigma_{j_0}=0.08$, $\sigma_{s_0}=1.95$, $\sigma_{c_0}=14.20$ and
$\sigma_{H_0}=0.48$, respectively. We hence see that the
constraints on the parameters provided by the future mock data can
be strongly improved in comparison with the current constraints in
Table \ref{table:4th}.

\section{Cosmographic selection of viable cosmological models}
\label{s6}

In the case of the standard flat $\Lambda$CDM model (namely a model
described by Cold Dark Matter with the adding of a cosmological
parameter) the set of cosmographic parameters results to be (up to
fifth order) \bea q_0&=&\frac{3}{2}\Omega_{m_0}-1\nn\\
j_0&=&1\nn\\
s_0&=&1-\frac{9}{2}\Omega_{m_0}\nn\\
c_0&=&1+3\Omega_{m_0}+\frac{27}{2}\Omega_{m_0}^{2}\;. \eea

We can use independent probes to constrain the free parameters of
the cosmological model, in this case, for example, the WMAP estimates of $\Omega_{m_0}$ for the
$\Lambda$CDM model.
%
%

The theoretical predictions of the cosmographic parameters in the
standard $\Lambda$CDM model are: $q_0=-0.588$, $j_0=1$, $s_0=-0.238$
and $c_0=2.846$, where we set the current matter density to be the
best fit value $\Omega_{m_0}=0.275$ \footnote{The estimate of
$\Omega_{m_0}$ is of course known within a certain error. From now
on, for illustrative purposes, we will retain the best fit values of
the free parameters, independently estimated in every single
cosmological model, as the fiducial ones, without taking into
account the associated errors.} obtained by the WMAP group
\cite{WMAP7}. Future experiments, in this perspective, will give
stricter constraints on the validity of such hypothesis.

Another example is provided by the Dvali-Gabadadze-Porrati (DGP)
self-accelerating brane\-world model \cite{dgp}. The presence of the
infinite-volume extra dimension modifies the Friedmann equation as:
\bea \frac{H^2}{H_0^2}=\Omega _{k }(1+z)^2+\left(\sqrt{\Omega
_{r_c}}+\sqrt{\Omega _{r_c}+\Omega _{m_0}(1+z)^3}\right)^2\,, \eea
with $\Omega _{r_c}=1/4 r_c^2 H_0^2$ accounting for the fractional
contribution of the bulk-induced term with respect to the crossover
radius $r_c$. In a spatially flat universe, $\Omega _{k }=0$ and
$\Omega _{r_c}={\left(1-\Omega _{m_0}\right)^2}/{4}$, the previous
equation reads \bea\label{dgp1}
\frac{H^2}{H_0^2}=\left[\frac{1-\Omega
_{m_0}}{2}+\sqrt{\frac{\left(1-\Omega _{m_0}\right)^2}{4}+\Omega
_{m_0}(1+z)^3}~\right]^2\,, \eea so, expanding both the sides of
Eq.(\ref{dgp1}) and equating terms of the same power, we obtains the
following expressions for the cosmographic coefficients as functions
of the free parameter $\Omega_{m_0}$ (see also Ref.\cite{DGPcosmo})
\bea q_0=\frac{-1+2 \Omega _{m_0}}{1+\Omega _{m_0}}\nonumber \eea
\bea j_0=\frac{1+3 \Omega _{m_0}-6 \Omega _{m_0}^2+10 \Omega
_{m_0}^3}{\left(1+\Omega _{m_0}\right){}^3}\nn \eea \bea
s_0=\frac{1-4 \Omega _{m_0}+19 \Omega _{m_0}^2-134 \Omega
_{m_0}^3+86 \Omega _{m_0}^4-80 \Omega _{m_0}^5}{\left(1+\Omega
_{m_0}\right){}^5}\nn \eea \bea c_0&=&\frac{1+13 \Omega _{m_0}\!-141
\Omega _{m_0}^2\!+1259\Omega_{m_0}^3 \!-1996 \Omega _{m_0}^4\!+3828
\Omega _{m_0}^5\!-1604 \Omega _{m_0}^6\!
+880 \Omega _{m_0}^7}{\left(1+\Omega _{m_0}\right){}^7}\,.\nn\\
\eea

In Ref.\cite{sereno}, the DGP model has been constrained starting
from gravitational lensing statistics; considering the fractional
amount of matter obtained therein, $\Omega_{m_0}=0.30$, we obtain
the following set of values for the previous parameters:
$q_0=-0.308$, $j_0=0.742$, $s_0=-0.432$, $c_0=2.926$.

We will now take into account the so-called Cardassian cosmology
\cite{card}, a model whose modification with respect to standard
$\Lambda$CDM cosmology consists in the introduction of an additional
term $\rho^n$ in the matter source of the Friedmann equation, so
that now it can be written in term of the fractional matter density
as: \be \frac{H^2}{H_0^2}=\Omega _{m_0}(1+z)^3+\left(1-\Omega
_{m_0}\right) (1+z)^{3 n}\,.  \ee

Performing one more time the expansion of both sides of the
equation, the first four cosmographic parameters can now be
expressed as functions of the two parameters $\Omega_{m_0}$ and $n$
\bea\label{carda}
q_0&=&\frac{1}{2}+\frac{3}{2} (1-n) \left(\Omega _{m_0}-1\right)\nn\\
j_0&=&\frac{1}{2} \left[2+9 n \left(\Omega _{m_0}-1\right)+9 n^2 \left(1-\Omega _{m_0}\right)\right]\nn\\
s_0&=&\frac{1}{4} \left[4-18 \Omega _{m_0}-9n \left(4-7 \Omega _{m_0}+3\Omega _{m_0}{}^2 \right)+9 n^2 \left(11-17 \Omega _{m_0}+6\Omega _{m_0}{}^2 \right)-\right.\nn\\
&&\left.-27 n^3 \left(3-4 \Omega _{m_0}+\Omega _{m_0}{}^2 \right)\right]\nn\\
c_0&=&\left(1+3 \Omega _{m_0}+\frac{117 \Omega _{m_0}^2}{2}-\frac{243 \Omega _{m_0}^3}{2}+\frac{1215 \Omega _{m_0}^4}{16}\right)-\nn\\
&&\hspace{0.5cm}-\frac{3}{4} \left(32-80 \Omega _{m_0}+291 \Omega _{m_0}^2-648 \Omega _{m_0}^3+405 \Omega _{m_0}^4\right) n+\nn\\
&&\hspace{1cm}+\frac{9}{8} \left(136-242 \Omega _{m_0}+349 \Omega _{m_0}^2-648 \Omega _{m_0}^3+405 \Omega _{m_0}^4\right) n^2-\nn\\
&&\hspace{1.5cm}-\frac{27}{4} \left(46-73 \Omega _{m_0}+54 \Omega _{m_0}^2-72 \Omega _{m_0}^3+45 \Omega _{m_0}^4\right) n^3+\nn\\
&&\hspace{2cm}+\frac{81}{16} \left(39-56 \Omega _{m_0}+26 \Omega
_{m_0}^2-24 \Omega _{m_0}^3+15 \Omega _{m_0}^4\right) n^4\,, \eea
and for the referring values $\Omega_{m_0}=0.271$ and $n=0.035$
\cite{card2}, Eq.(\ref{carda}) reads $q_0=-0.555$, $j_0=0.890$,
$s_0=-0.384$, $c_0=3.660$.

Finally, we want to show the coefficients in the cosmographic
approach of the CPL parametrization \cite{cpl} for the equation of
state of Dark Energy. If we suppose to be in a flat universe, then
the Friedmann equation is: \be \frac{H^2}{H_0^2}=\Omega
_{m_0}(1+z)^3+\left(1-\Omega _{m_0}\right) (1+z)^{3
\left(1+w_0+w_a\right)} e^{-\frac{3 w_a z}{1+z}}\,, \ee and the
related cosmographic terms result to be (confront also with ref.
\cite{luca}) \bea
q_0&=& 1+\frac{3}{2} w_0 \left(1-\Omega _{m_0}\right)\nn\\
j_0&=& 1+\frac{3}{2} \left(3 w_0+3 w_0^2+w_a\right) \left(1-\Omega _{m_0}\right)\nn\\
s_0&=&-\frac{7}{2}-\frac{33}{4} \left(1-\Omega _{m_0}\right) w_a-\frac{9}{4}\left(1-\Omega _{m_0}\right) \left[9+\left(7-\Omega _{m_0}\right) w_a\right] w_0-\nn\\
&&-\frac{9}{4}\left(1-\Omega _{m_0}\right) \left(16-3 \Omega _{m_0}\right) w_0^2 -\frac{27 }{4}\left(1-\Omega _{m_0}\right) \left(3- \Omega _{m_0}\right) w_0^3\nn\\
c_0&=&\frac{1}{4} \left(70+3 w_a \left(-71+3 w_a \left(-7+\Omega _{m_0}\right)\right) \left(-1+\Omega _{m_0}\right)\right)+\nn\\
&&+\frac{3}{4} \left(-1+\Omega _{m_0}\right) \left(-163+3 w_a \left(-82+21 \Omega _{m_0}\right)\right) w_0+\nn\\
&&+\frac{9}{4} \left(-1+\Omega _{m_0}\right) \left(-134-69 w_a+3 \left(14+11 w_a\right) \Omega _{m_0}\right) w_0^2+\nn\\
&&+\frac{1}{4} \left(1269-1917 \Omega _{m_0}+648 \Omega _{m_0}^2\right) w_0^3+\frac{1}{4} \left(486-810 \Omega _{m_0}+324 \Omega _{m_0}^2\right) w_0^4\,;\nn\\
\eea assuming the values suggested by the seventh-year-release of
WMAP \cite{WMAP7} for the three free parameters,
$\Omega_{m_0}=0.275$, $w_0=-0.93$ and $w_a=-0.41$, we get
$q_0=-0.511$, $j_0=0.342$, $s_0=-2.260$, $c_0=1.383$. Table
\ref{confr} shows the values of the cosmographic parameters in the
different models taken into account.
%

It is interesting to note a couple of issues in the comparison of
cosmological models with our best fits. Firstly, the best fit for
the cosmographic parameters of the SN+GRB+BAO+CMB data set is
perfectly compatible with the estimates of the cosmological
parameters for a broad variety of models. However, it is easy to
realize that the currently available data sets would not allow yet
to distinguish among the different cosmological models. In fact,
Table \ref{confr} shows that the error bars are still too large with
respect to the differences among the cosmographic parameters of the
cosmological models. Nonetheless, the previously discussed
forecasted improvement in the quality and the quantity of such data
(ameliorating by at least a factor ten the error bars on the
cosmographic parameter) should be able to discriminate among
competing models.

On the contrary, the best fit of the widest combination of data
(that is with the inclusion of the Hubble parameter determinations
via the differential age technique), seems to exclude, at a
3-$\sigma$ around the jerk mean value $j_0$, almost all cosmological
models, including $\Lambda$CDM (with the only exception of the CPL
modelization, that is still marginally compatible). We have already
discussed the intrinsic difference of the Hubble data and why their
use should be taken cautiously. It seems clear that this data set
while being very powerful in reducing the error bars, is
simultaneously introducing strong deviations from the mean values
determined via standard candles and rulers. This puzzling
discrepancy in the results does not seem related to the order of the
truncation: we observed a similar behavior even for (statistically
not favored) early or late truncations of the series.

However, it is also true that the high redshift measurements of the
Hubble parameter are based on fitting galaxy spectra. As such, this
data set is strongly dependent on this fitting procedure which may
introduce systematic effects. For this reason, we deem estimates
based on the Hubble data currently less robust than those based on
standard candles and rulers. Nonetheless, their use here serves to
show the possible key role these data could play in the future of
Cosmography as they appear to be very effective in reducing the
error bars and very sensitive tracers of the cosmological history.
We hence conclude that our analysis strongly suggests further
investigation of this apparent tension between the Hubble data and
$\Lambda$CDM (and many competing models) via a refinement of the
determination methods developed in Refs.\cite{Jimenez:2003iv,
Simon:2004tf}.

\begin{table*}
\caption{Comparison among cosmographic parameters of different
cosmological models. For every model, the evaluation of the
cosmographic parameters, for a pedagogical issue, is based on the
best fit values of the free parameters introduced in the dynamics
and measured with independent probes. However, the value of the jerk
parameter for $\Lambda$CDM model is an exact value, see also (17).
The values of the cosmographic parameters are compared with our best
fits for the series truncations studied in the last two lines of
Table I. The distances indicators data set includes SNIa, GRB, BAO,
CMB. The complete data set is obtained adding Hubble
data.}\label{confr}
\begin{center}
\begin{tabular}{c|cccc}
  \hline
  \hline
  Parameter&$q_0$&$j_0$&$s_0$&$c_0$\\
  \hline
  $\Lambda$CDM&$-0.588$&$1$&$-0.238$&$2.846$\\
  DGP&$-0.308$&$0.742$&$-0.432$&$2.926$\\
  Cardassian&$-0.555$&$0.890$&$-0.384$&$3.660$\\
CPL Parametrization&$-0.511$&$0.342$&$-2.260$&$1.383$\\
  \hline
  \hline
Best fit &&&&\\
SN+GRB+BAO+CMB ($5^{th}$ order)&$-0.49\pm0.29$&$-0.50\pm4.74$&$-9.31\pm42.96$&$126.67\pm190.15$\\
SN+GRB+BAO+CMB ($5^{th}$ order) + Hub ($4^{th}$ order)&$-0.30\pm0.16$&$-4.62\pm1.74$&$-41.05\pm20.90$&$-3.50\pm105.37$\\
\hline
\end{tabular}
\end{center}
\end{table*}

\section{Discussion and conclusion}
\label{s7}

Reaching the highest possible redshift allowed by data is a
fundamental tool to discriminate among competing cosmological
models. Given that most of the models are built in order to recover
Dark Energy at low redshift, their expansion histories are
degenerate at late times. To break such a degeneracy, it is required
an improvement on the knowledge of the early universe expansion
curve: this aim can be achieved only by an accurate determination of
the higher order parameters, and higher terms in the cosmographic
expansion can be consistently reached only using (very) high
redshift data.

Compared to our previous work we have added Baryonic Acoustic
Oscillations (that are distance indicators at $z\sim 0.3$), a
compilation of high redshift Hubble parameter measurements and, at
least for a wide gamut of cosmological models, CMB data about the
angular size of the sound horizon. This improved data set is helpful
in that, apart from better constraining the previously studied
cosmographic parameters, it also allows to cast constraints on the
next order, so far unbound, expansion coefficient.

The analysis is performed by using Monte Carlo Markov Chains in the
multidimensional parameter space to derive the likelihood. As a
first step, we consider the most recent catalogs of standard
candles, namely Supernovae Type Ia and (properly standardized, see
discussion in Section \ref{s3}) GRBs. A combination of such data
gives constraints up to the $s_0$ parameter in the cosmographic
series. We have also used the BAO (albeit the mildly improve the
cosmographic series fitting) discussing the reliability of such
tools in this context.

Secondly, we add data at higher redshift from different probes to
further improve the constraints. The CMB data account for a very
stable and well determined scale. It is worth noting here, anyway,
that on the contrary of the other probes, CMB data provide the
problem of a lack of universality in the cosmographic approach.
Unfortunately, the set of parameters extracts from CMB observations
is not truly independent from the dynamics of the underlying
gravitational theory. Its definition, in fact, strictly depends on
the assumption of a cosmological model that behaves as General
Relativity plus a content of matter of arbitrary nature. It is hence
impossible to use it straightforwardly within a purely cosmographic
analysis which wants to apply also to non-standard cosmologies
(based on exotic modified gravity theories)\footnote{Of course, CMB
observables can be used within a given gravitational dynamics to fix
the free variables of a cosmological model \cite{elga} and hence
calculate the corresponding cosmographic parameters to be confronted
with those determined purely on the base of standard candles and
rulers, as we also showed as application to the evaluation of
cosmographic parameters in several cosmological models.}. In this
paper we proposed CMB data constraints on the cosmographic series by
restricting the results to a slightly smaller variety of models. A
desirable full solution to this problem can be achieved
``standardizing'' somehow the CMB parameters or alternatively
identifying other CMB observables which could be used as standard
rulers (at least approximately, as for BAOs). We leave this to
future investigations.

We then added the high redshift measurements of the Hubble
parameter. We found that thanks to these data and the CMB one, it is
possible to ameliorate the knowledge of the cosmographic expansion
up to the $c_0$ parameter.

As a completion of our analysis, we have also discussed foreseeable
constraints from futuristic data sets provided by projected
experiments. We showed that a strong reduction of the typical errors
on the parameters estimates is a realistic goal: future surveys,
indeed, do have a solid possibility to sufficiently reduce the
uncertainties on the lowest order parameters by a factor ten at
least, gaining a concrete chance to assess the viability of
alternative cosmological models (possibly based on different
dynamics).

Finally, we calculated the cosmographic parameter sets for a sample
of cosmological models with alternative dynamics (using the so far
available best fits for their free parameters). We showed that,
while the data set including ``standard'' distance indicators gives
a best fit with which all the cosmological models are still
compatible, the inclusion of the Hubble data introduces a tension
between the observed cosmographic parameters and the parameters
calculated for different models and in particular with $\Lambda$CDM
which appears to be ruled out at 3-$\sigma$ due to the jerk best fit
value. We have discussed the reliability of such observation taking
into account the inhomogeneity of the Hubble data set with respect
to the distance indicators ones. While there might be systematic
uncertainties in this data due to their complex determination
method, we stressed that our analysis strongly suggests they might
play a key role in reducing the errors on the estimates of the
cosmographic parameters and hence in making Cosmography effective in
discriminating among competing cosmological models and gravitational
theories.

In conclusion, the search for high redshift standard rulers and most
of all the improvement of the data coming from galaxy surveys seem
to be what could possibly bring cosmographic studies into a mature
stage and make them powerful, gravity theory independent, tools for
selecting among theoretical scenarios. We hence hope that this work
will further strengthen the case for proposed experiments aimed at
improving our knowledge of the cosmic evolution of the high redshift
universe (e.g. Ref.\cite{Liske}).

\section*{Appendix: Series Expansions}

We present here more extensively the expansions used in this work. A
flat universe, $k=0$, is assumed in all the expressions below.\\
Hubble parameter: \bea
H[z(y)]=H_0&&\Bigg[1+(q_0+1) y+y^2 \left(\frac{j_0}{2}-\frac{q_0^2}{2}+q_0+1\right)+\nn\\
&&+\frac{1}{6} y^3 \left(-4 j_0 q_0+3 j_0+3 q_0^3-3 q_0^2+6
   q_0-s_0+6\right)+\nn\\
&&+\frac{1}{24} y^4 \left(-4 j_0^2+25 j_0 q_0^2-16 j_0 q_0+12 j_0+c_0-15 q_0^4+\right.\nn\\
&&\left.\hspace{3cm}+12 q_0^3-12 q_0^2+7 q_0
   s_0+24 q_0-4 s_0+24\right)\Bigg]+\mathcal{O}[y^5]\,;\nn\\
\eea
Luminosity distance: \bea
d_L[z(y)]=\frac{1}{H_0}&&\Bigg[y+\left(\frac{3}{2}-\frac{q_0}{2}\right)
y^2 +y^3 \left(\frac{q_0^2}{2}-\frac{5
   q_0}{6}-\frac{j_0}{6}+\frac{11}{6}\right)+\nn\\
&&+y^4 \left(\frac{5 j_0 q_0}{12}-\frac{7
   j_0}{24}-\frac{5 q_0^3}{8}+\frac{7 q_0^2}{8}-\frac{13
   q_0}{12}+\frac{s_0}{24}+\frac{25}{12}\right)+\nn\\
&&+y^5
   \left(\frac{j_0^2}{12}-\frac{7 j_0 q_0^2}{8}+\frac{3 j_0
   q_0}{4}-\frac{47 j_0}{120}-\frac{c_0}{120}+\frac{7 q_0^4}{8}-\frac{9
   q_0^3}{8}+\frac{47 q_0^2}{40}-\right.\nn\\
&&\left.\hspace{5cm}-\frac{q_0 s_0}{8}-\frac{77
   q_0}{60}+\frac{3 s_0}{40}+\frac{137}{60}\right)\Bigg]+\mathcal{O}[y^6]\,;\nn\\
\eea
Angular distance: \bea
d_A[z(y)]=\frac{1}{H_0}&&\Bigg[y-\left(\frac{q_0}{2}+\frac{1}{2}\right) y^2+y^3 \left(\frac{q_0^2}{2}+\frac{q_0}{6}-\frac{j_0}{6}-\frac{1}{6}\right)+\nn\\
&&+y^4 \left(\frac{5 j_0 q_0}{12}+\frac{j_0}{24}-\frac{5
   q_0^3}{8}-\frac{q_0^2}{8}+\frac{q_0}{12}+\frac{s_0}{24}-\frac{1}{12}\right)+\nn\\
&&+y^5 \left(\frac{j_0^2}{12}-\frac{c_0}{120}-\frac{7 j_0
q_0^2}{8}-\frac{j_0
q_0}{12}+\frac{j_0}{40}+\frac{7q_0^4}{8}+\frac{q_0^3}{8}-\frac{3
   q_0^2}{40}-\frac{q_0 s_0}{8}+\right.\nn\\
&&\left.\hspace{6cm}+\frac{q_0}{20}-\frac{s_0}{120}-\frac{1}{20}\right)\Bigg]+\mathcal{O}[y^6]\,;\nn\\
\eea
Volume distance: \bea
d_V[z(y)]=\frac{1}{H_0}&&\Bigg[y+\left(\frac{1}{3}-\frac{2
q_0}{3}\right) y^2+y^3 \left(\frac{29
   q_0^2}{36}-\frac{5 q_0}{18}-\frac{5 j_0}{18}+\frac{7}{36}\right)+\nn\\&&+y^4 \left(\frac{43 j_0 q_0}{54}-\frac{13 j_0}{108}-\frac{94 q_0^3}{81}+\frac{13 q_0^2}{36}-\frac{19
   q_0}{108}+\frac{s_0}{12}+\frac{11}{81}\right)+\nn\\
&&+y^5\! \left(\!\frac{59 j_0^2}{324}-\frac{605 j_0
q_0^2}{324}+\frac{233 j_0 q_0}{648}-\frac{32 j_0}{405}-\frac{7
   c_0}{360}+\frac{7079 q_0^4}{3888}-\frac{523 q_0^3}{972}\right.+\nn\\ &&\left.\hspace{2cm}+\frac{773 q_0^2}{3240}-\frac{5 q_0 s_0}{18}-\frac{623 q_0}{4860}+\frac{13
   s_0}{360}+\frac{2017}{19440}\right)\!\!\Bigg]\!+\mathcal{O}[y^6]\,.\nn\\
\eea

\section*{Acknowledgments}

The authors are really grateful to the referee for crucial
suggestions to improve the paper and to B.~Bassett and M.~Visser for
their useful comments. VV wishes to thank warmly L.~Izzo and
M.~Martinelli for fruitful discussions. MV is supported by
PRIN-MIUR, PRIN-INAF 2009, INFN/PD-51, ASI/AAE grant and the
European ERC-StG ``cosmoIGM'' grant.


\end{document}